# A NEW PARADIGM IN SPACE BASED EXPERIMENTS USING RUBBER BALLOONS

Sandip K. Chakrabarti[2,1], Debashis Bhowmick[1], Sourav Palit[1], Subhankar Chakraborty[1], Sushanta Mondal[1], Arnab Bhattacharyya[1], Susanta Middya[1], Sonali Chakrabarti[1]

[1]*Indian Center for Space Physics, 43 Chalantika, Garia Station Rd., Kolkata 700084, INDIA*
[2]*S.N. Bose National National Centre for Basic Sciences, JD Block, Salt Lake, Kolkata, INDIA*
Email:chakraba@bose.res.in

## ABSTRACT

Indian Centre for Space Physics is engaged in long duration balloon borne experiments with typical payloads less than ~ 3kg. Low cost rubber balloons are used. In a double balloon system, the booster balloon lifts the orbiter balloon to its cruising altitude where data is taken for a long time. Here we present results of muon detections and recent solar activities, including the light curves and flare spectra in the 20-100keV range. We not only show that we have successfully obtained several flares and there spectra at different altitudes, we also found that the high energy X-ray flux of strong flares at altitudes of 10-13 km (the flight altitude of commercial planes) could be more than the contribution due to cosmic rays.

## 1. INTRODUCTION

Indian Centre for Space Physics is interested in carrying out balloon borne experiments at a sustainable cost so that going to near space for quality data is possible on a daily basis. With miniaturization of instruments, such a goal appears to be attainable. Here we present some results of obtaining high quality data with payloads having a total mass of 2-3 kg or less. Instrumentations are fabricated in house. All payloads are returned by parachutes. In Chakrabarti et al. (2011) our early studies have been reported.

## 1. TYPICAL LAUNCH PROFILES

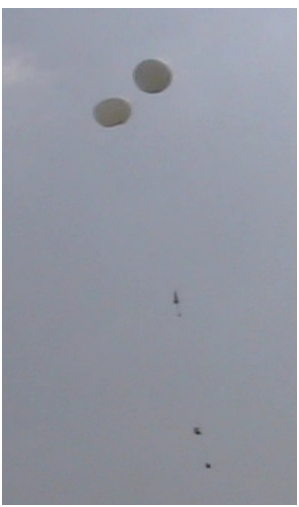

*Figure 1. A double balloon being launched with a single parachute, communication box and the payload.*

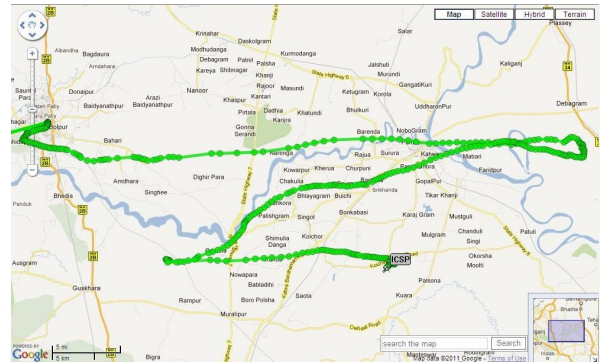

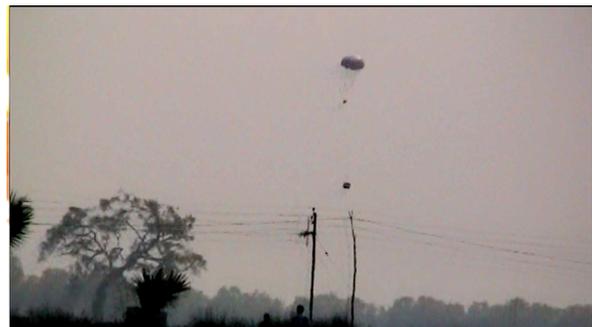

*Figure 2. A typical trajectory of a balloon from launch site to landing site (top); The payload is returned by the parachute (bottom).*

ICSP launches from Bolpur (23.67N, 87.72E). However, since we launch small balloons (Figure 1), no launch `facility' is necessary and often we choose the locations according to where we wish the parachute dropping of the payload (Figure 2). Our location on the Tropics of Cancer compensates for the Earth's tilt in summer and thus solar data is obtained without any special pointing equipment. Typically, every launch contains GPS data unit, GPS tracker, 9DOF, camera, parachute(s), communication box, sun-sensor, power supply and the Payload. We have developed Sun-tracker and it would be used in future missions when we shall observe other objects near galactic centre in relation to the sun.

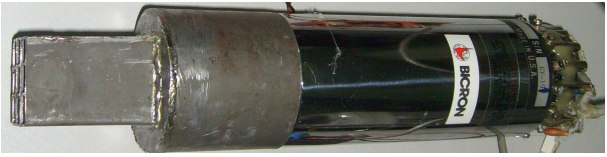

*Figure 3. Bicron gamma-ray detector (2 inch diameter) with a lead shielding. This instrument has been launched in nine Dignity missions.*

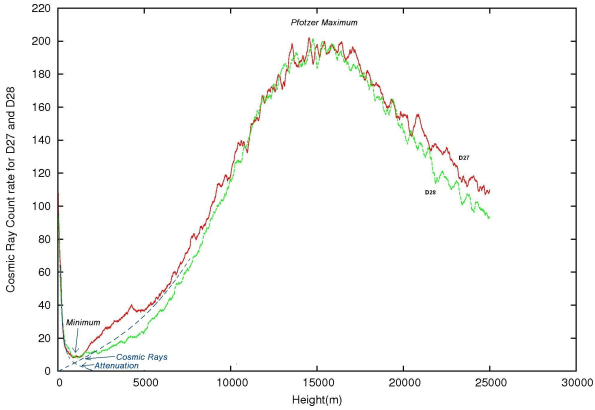

*Figure 4. Examples of cosmic ray profiles (60 second average) in Dignity 27 and 28 missions by a bicron detector having a PMT and a NaI crystal. At our latitude the Pfotzer maximum is at ~15 km. The minimum caused by the competition of ground radio activity and Cosmic rays is marked.*

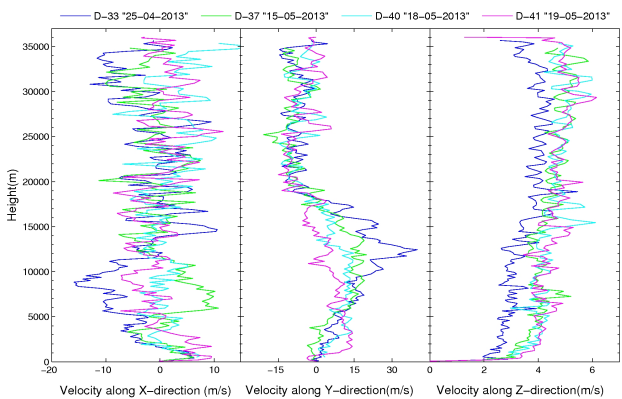

*Figure 5. Velocity components at our latitude in pre-monsoon season in Dignity missions 33 (D33), 37 (D37), 40 (D40) and 41 (D41).*

The same payloads are repaired, if needed, and are launched again on the next day if wind condition is favourable. The typical non-recoverable cost is about $500 per Mission if no new payload is attached.

In Figure 3 we present the photo of a payload instrument (Bicron made gamma-ray detector, 2 inch diameter), with a 0.5mm lead shielding to block hard X-rays below 100 keV. The collimator emits a line at 77keV which is used as a calibrator. All the detectors are calibrated on the ground during testing, and before and after the launching. In Figure 4, we show examples of cosmic rays we received from the bicron detector on two successive days (24/5/2012 and 25/5/2012). At our latitude we consistently have the maximum at around 15-16km. There is a distinct minimum just after launch, due to the attenuation of the ground radio-activity and the rise in cosmic rays with height. In Figure 5, we show the wind profiles measured by our GPS system. We superposed four days of data in the pre-monsoon season, which clearly indicate the switching of X- and Y- components of velocity. Typical Z-component of velocity is ~ 4 m/s.

## 2. LONG DURATION STUDIES WITHOUT VALVES OR BALLAST

To achieve long duration we have been using two balloons, one being the booster balloon with a larger lift while the other is an orbiter balloon with a smaller lift. The Booster bursts (Fig. 6) at its burst height (say, 37km), and the orbiter cruises at a desired altitude of 25-33km depending on initial lift and payload weight.

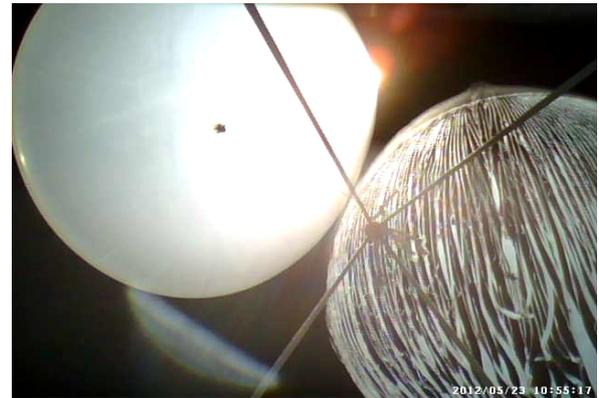

*Figure 6. A double balloon with the booster bursting as seen from the payload camera. The sun is behind the orbiter balloon which is in tact.*

The theory of placing a balloon at the cruising orbit is simple yet very robust (Chakrabarti et al. 2013) and will not be repeated here. It is easily shown that assuming (a) $P_i = P_o+P_w$ (i and o stand for internal and external to the balloon; w stands for balloon wall) (b) $T_i=T_o$, (c) ideal gas law $P_iV_b=nRT_i$ (d) Mooney-Rivlin model of hyperelastic material (Mooney, 1940; Rivlin, 1948). P,V,T, and R have usual meaning. The balloon characteristics is vary important to achieve this.

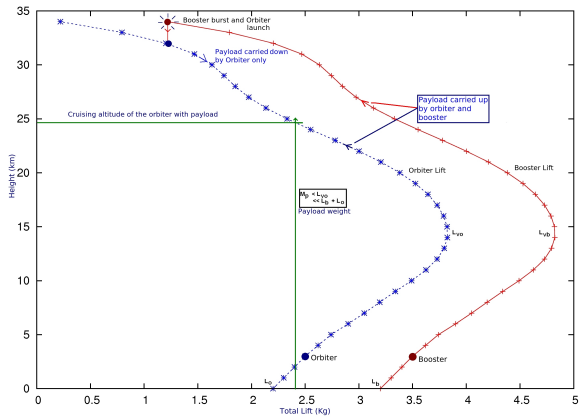

*Figure 7. Analytical solution of a double balloon configuration with the Booster bursting at 40km and launching the orbiter at 25km. The region of the curve with the dHeight/dLift<0 is stable for long cruising.*

In Figure 7, we demonstrate the principle of how we achieve long duration (anywhere between 5 to 20 hours) without using any valve or ballast with an example where the booster and the orbiter balloons are given lifts of 2.2kg and 3.2kg respectively while the payload is of weight 2.4kg. Initially lifts increase (we call it a *lift valve*, as it helps during ascent, but slows down the descent). The Booster bursts at its burst height due to higher lift, while the orbiter slowly descends down to its neutral buoyancy altitude and orbits there till it is ejected off the orbit or naturally drops due to cooler atmosphere at late night. In Figure

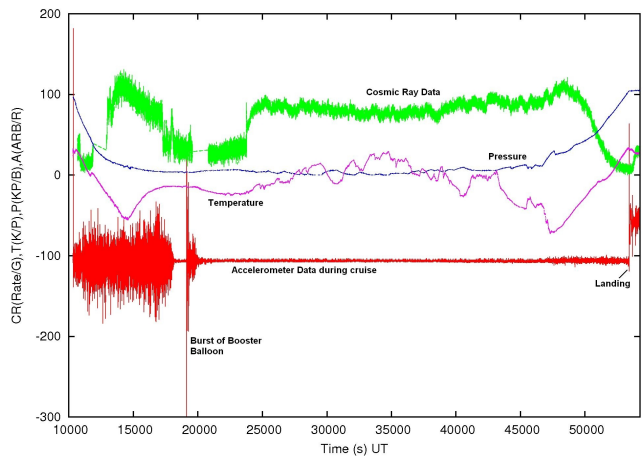

*Figure 8. The cosmic rays (Green), pressure (blue), temperature (pink) and the Accelerometer (red) data for a long duration flight (Dignity 26) which lasted 12 hours. The booster burst at 37.9km and the orbiter was cruising at 25km. There were clear indications (temperature data) of some oscillations of height. Accelerometer data indicates that the flight was smooth.*

8, we show an example of the results by a long duration flight where a Hamamatsu made photo-multiplier tube with NaI crystal was chosen as the X-ray detector (Dignity 26 mission on 23/5/2012, lift off at 3:01UT from 22.37N,88.446E, Landing after 11h58m, at 23.96N,84.65E) and camera, 6DOF, parachutes are attached.

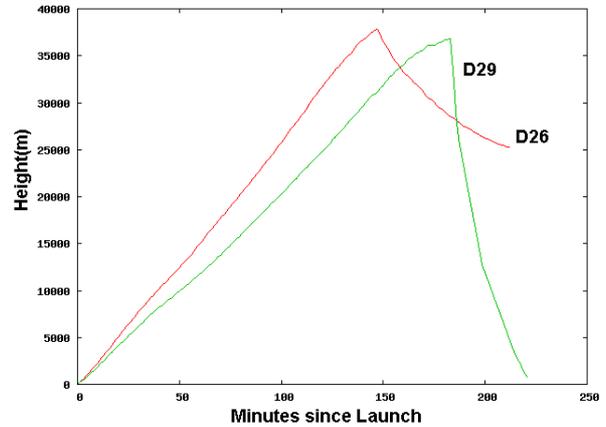

*Figure 9. Time-Height profiles of two cases: Dignity-29 (D29) had no orbiter, Dignity-26 (D26) had an orbiter which settled in an orbit at ~ 25 km after initial fluctuations are settled down.*

The characteristics of the descent of the orbiter after the burst of the booster is totally different. In Figure 9, we show that in Dignity-29 where there was no orbiter, the descent was normal, in about 45 minutes, by the parachute. However, the payload in Dignity-26 took long time to descend.

## 2. STUDIES OF SOLAR FLARES

Since 24th Solar cycle is approaching, ICSP engaged in constructing payloads with a goal to observe solar flares. In 2012, when the solar activities were lower, we failed to obtain many flares, though we obtained the spectra of the quiet sun. Our lead-shielding (of 0.5mm) could not be made thicker due to weight constraints. The collimator, also made by lead was of 40º x 40º to begin with, to study the proof of concept. To enable us to view the sun for a maximum amount of time, we adjusted the tilt of the payload with respect to the zenith so that the sun is close to the centre of the collimator when the balloon is at a high altitude. In any case, our 9 degrees of freedom (9DOF) chips enabled us to determine the RA & DEC of the axis of the payload post facto, to verify which direction we were pointing at. The on-board video camera is used to measure the balloon diameter and also the burst characteristics. The sun-sensor stamps every frame as to whether or not the sun was inside the collimator. All these ensure us that were indeed observing the sun. In Figure 10, we present all sky location of the bright X-ray/gamma-ray sources from SWIFT/BAT catalogue. The location of Sun on 18th May, 2013 is shown. The sun is surrounded by a grey area which is covered by the axis of the collimator in five minutes before the burst (Dignity 40). The red curve indicates the coverage by a corner of the collimator. A post-facto picture of this kind, along with the on-board video camera and sun-sensor give us the time stamps of the data when our instrument was pointing to the Sun,

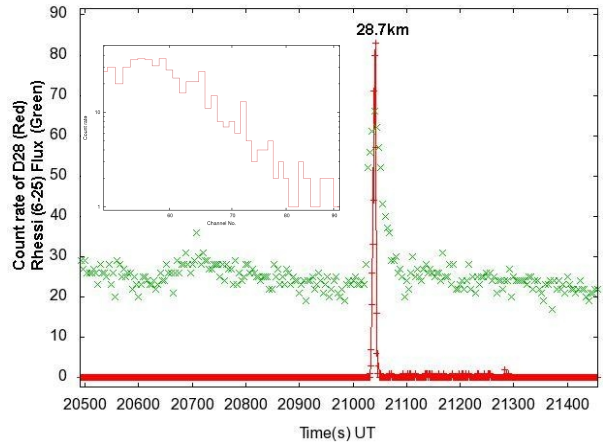

*Figure 11. A short duration flare observed in D28 mission. Count rates in 50-100 channels (~ 20-25keV) is plotted. For comparison we superposed RHESSI flux data in 6-25 keV range. In the inset, we show the flare spectrum in log-log scale which is clearly a power-law spectrum. There is some absorption due to atmosphere in the lower energy.*

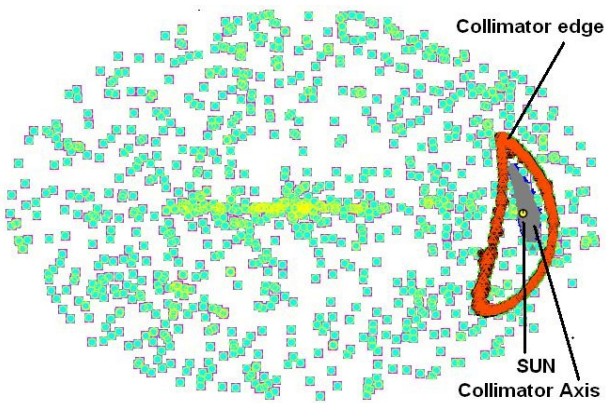

*Figure 10. All sky bright X-ray/gamma-ray sources from SWIFT/BAT catalogue is super-imposed with RA/DEC information from the 9DOF sensors on board.*

First, we show a weak solar flare we observed in 2012 in Dignity 28 mission (Fig. 11). All the data presented in this paper are from Bicron detector (Fig. 3). D27-D30, D33, D36-D37, D40-D41 missions carried this detector. Our light curve (time vs. Photon count rate) in D28 mission is compared with RHESSI Flux variation. Since RHESSI data included softer X-rays, its flare is wider. The inset shows the spectrum.

As the solar activities increased in 2013, we detected several solar flares. In Figure 12. we present the light curves (3s average of the raw data) of three solar flares which are observed when the balloon was at heights of 25km, 28km and 32-34km respectively. Clearly we anticipate that at lower altitude the spectrum would be harder, and as the height goes up, the spectrum would be softer and more intense. This is precisely what we see. Figure 13 shows the channel wise spectra for these flares (whole first flare data and 60s data at the rising phase of the other two flares). In Figure 14, we show the spectra of the same flares after calibration. We find that the inner most flare observed at 25km extends to harder X-rays and the detected energies of the flares progressively become softer and brighter. All these flares match with GOES and RHESSI data in time and shape.

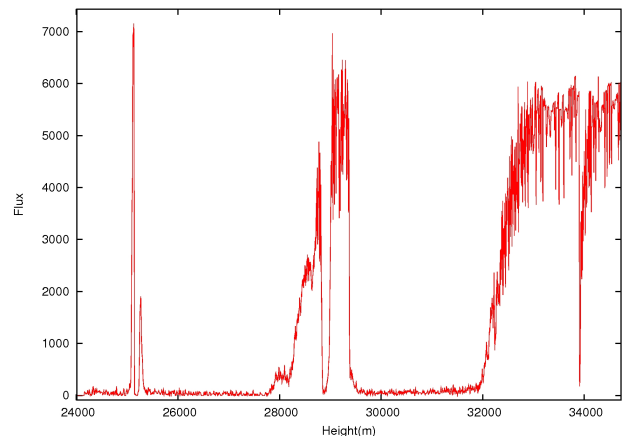

*Figure 12: Raw light curves of the three flares observed in D33 mission (25 Apr, 2013).*

In Figure 15, we show the dynamic spectrum (averaged over 60seconds, and binned in 1keV) of these flares. This diagram also clearly indicates the progressing increase in the detection of softer photons as the balloon climbs to higher altitude. In future these observed spectra would be reproduced through GEANT4 simulations incorporating the absorption by the atmosphere.

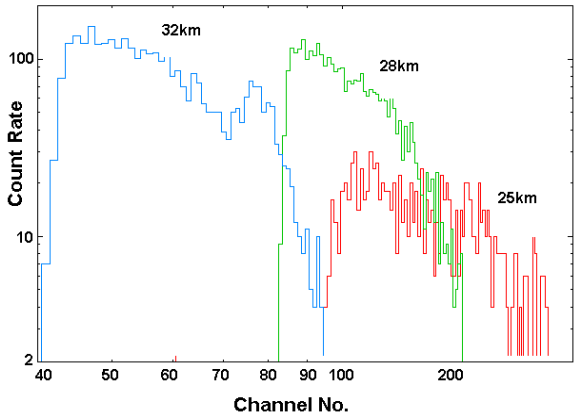

*Figure 13: Spectra of the rising phase of the three flares (Channel vs. Photon count rates) for the same three flares shown in Figure 12.*

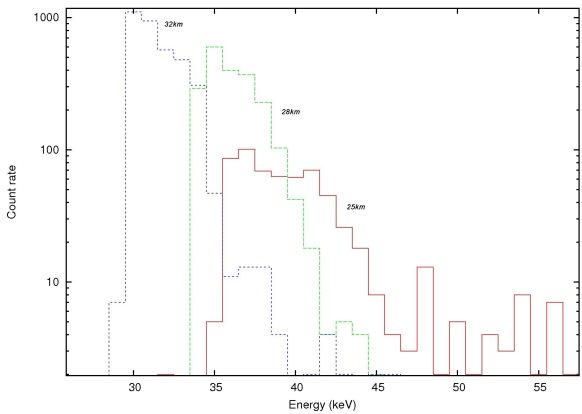

*Figure 14: Energy vs. Photon count rates of the rising phase of the three spectra as in Figures 12 and 13.*

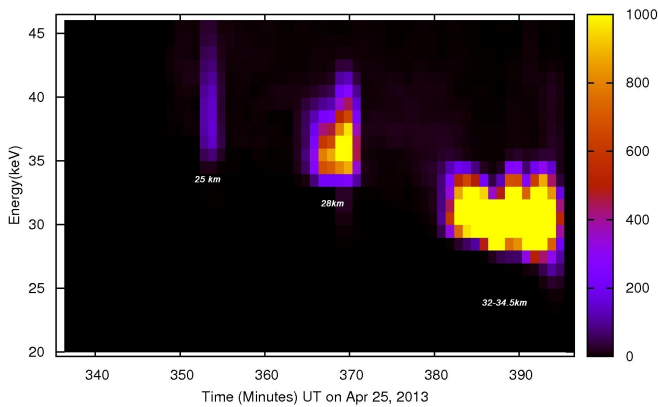

*Figure 15: Dynamical spectra of the observed data, showing clearly the flares and their time-resolved spectra. Note the progressive softening of the detection limit of the spectra with time.*

We now turn our attention to a strong flare which took place on 15th May, 2013. At ~ 1:30UT an X-class flare occurred and by the time we launched the balloon at 3:00UT, the flux has become M1. We detected excess X-rays right after the launch from ~ 8-9km till the end of burst of the balloon at 34.9km. In Figure 16, we show the count rates (20sq cm. Bicron detector) as a function of the height and channel number up to 17km in order to show details. We clearly see the effects of the flare in the upper branch marked as solar flare. The high counts in lower channels are from Cosmic rays. Surprisingly, the count rate at ~12-13km, the cruising altitudes of the commercial aircrafts is significant and energy wise comparable or more than that of the cosmic ray contribution. Judging from what is received at a height of 12km for a C4 flare (which is 25 times weaker than a X1flare that started at 1:30UT) when out instruments were at that altitude, we can conclude that the commercial flights are vulnerable to very high dose of radiation during strong solar flares. For instance those flying at 1:30UT were receiving at least 25 times more radiation than normal cosmic rays contribution at that altitude. Note that there is a minimum in high energy photons and very high number of low energy photons at Pfotzer maximum (~15km). The exact reason would be studied through GEANT4 simulation.

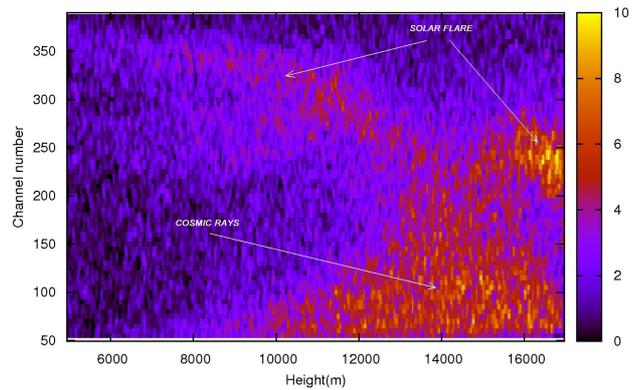

*Figure 16: Alarming rate of high energy photons detected at lower altitudes due to an X1 flare occurred at 1:30 UT, 15th May, 2013. When the instruments were passing through 12km, the flare was only C4, but the radiation dose was comparable to that from cosmic rays at that height. Count rate is for 20 sq. cm Bicron.*

To compare with an ordinary day, where there was no solar activity, we present in Figure 17, the dynamic spectrum obtained by our D29 mission (4 June, 2012). Clearly, we see only the cosmic ray component. It is not yet clear whether we are watching the direct Solar flare component, or the interaction of solar cosmic rays with the atmosphere. The details are being studied through GEANT4 simulations.

If we now turn to the main flare detection, we find that after the air becomes transparent enough, the entire flare could be detected. In Figure 18, we show the GOES 15 Satellite light curve (arbitrary unit) and our

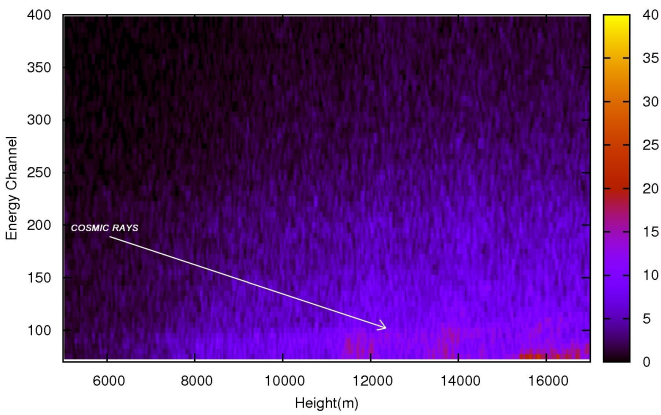

*Figure 17: Only cosmic rays were detected on D29 mission on 4th June, 2012.*

D37 mission flux in the 21-84keV range (Energy in each channel X count rate). After Pfotzer maximum our light curve started matching with the Satellite data and that too for high energy channels.

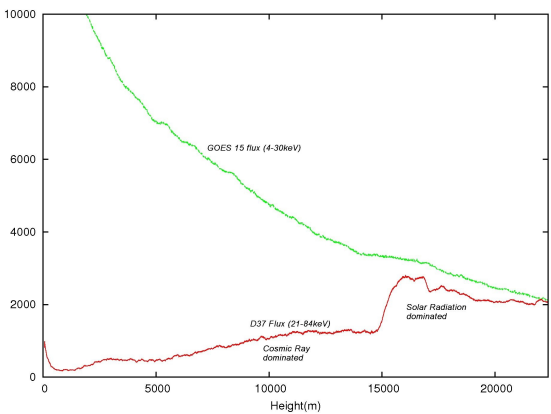

*Figure 18: Light curve of 21-84keV energy flux averaged over 60s is compared with GOES 15 4-30keV energy light curve. The difference decreases as the height increases.*

In order to find out how the instruments behave in a Styrofoam box in our payload, we sent a calibrator (apart from the lead shielding which was emitting at 77keV in all our missions), namely Eu152. We find that the energies drift upward (i.e., same energy appearing at a higher channel) as the instrument reaches the temperature inversion layer. After that it slowly returns back to normal. The internal temperature went down from room temperature to about 0 degree Celsius. This allowed us to calibrate the instruments dynamically. We also find that at lower temperature the energy resolution becomes excellent, only about 2keV FWHM at 39keV. In Figure 19, we show an example of the dynamically calibrated data from Bicron detector in Dignity 40 mission (18th May, 2013). We also observe a remaining of a weaker solar flare (C1), just after Pfotzer maximum.

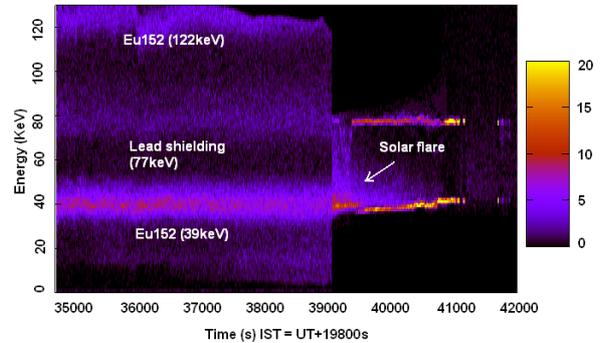

*Figure 19: Calibrated dynamical spectrum showing the Eu152 lines (39keV and 122keV) and the line due to lead shielding (77keV). We see the remnant of a C1 solar flare (see, Figure 20 for details).*

Since the data is fairly contaminated by the calibrators at 39keV and 77keV, we show in Figure 20 the dynamical spectrum of the solar flare between 42 and 72 keV. We find that, though the flare itself started when the instrument was at 14000m, we observe very prominently only after the resolution of the instruments became better. This was also the region after the Pfotzer maximum. Data is binned in one second in both the Figures 19 and 20.

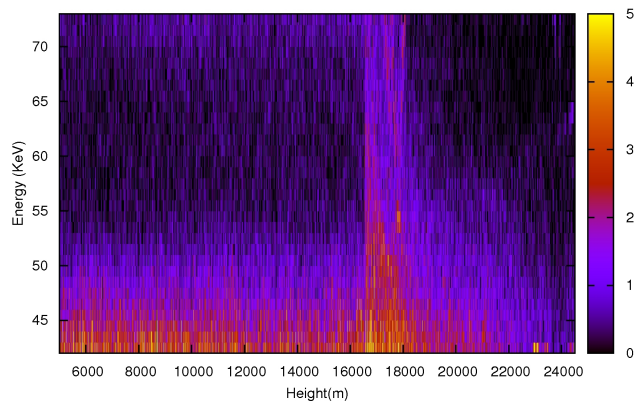

*Figure 20: Dynamical spectrum of a weak solar flare as a function of height as observed by D40 mission on 18th May, 2013. The data is plotted for 42keV to 72keV to avoid contaminations from the calibrators.*

In Figure 21, we show the actual spectra at three different heights, 19.5km, 20.9km and 22km which were achieved at a five minute interval. The spectra are obtained with one minute data. The plot is shown from 40 to 80keV.

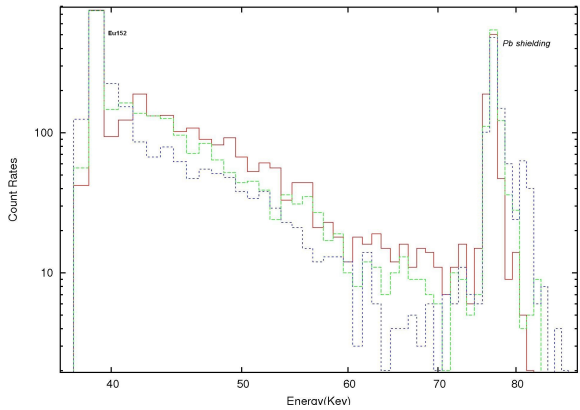

*Figure 21: Spectra of the C1 flare at three different heights when the instruments were at 19.5km (red), 20.9km (green) and 22.1km (blue) respectively. The calibrator lines which are very narrow and could be modelled and eliminated, here left out for illustration. The power-law nature of the spectra is obvious.*

## 3. OTHER STUDIES

ICSP is engaged in Muon detections, collection of micrometeorite samples and other developmental studies to have longer duration balloons. We present here a few examples.

**Muon Detection:** In Figure 22, we present the lead shielding (1-1.5cm thick) used for our muon detection. We used small 40.8mm long GM counters inside these lead shields.

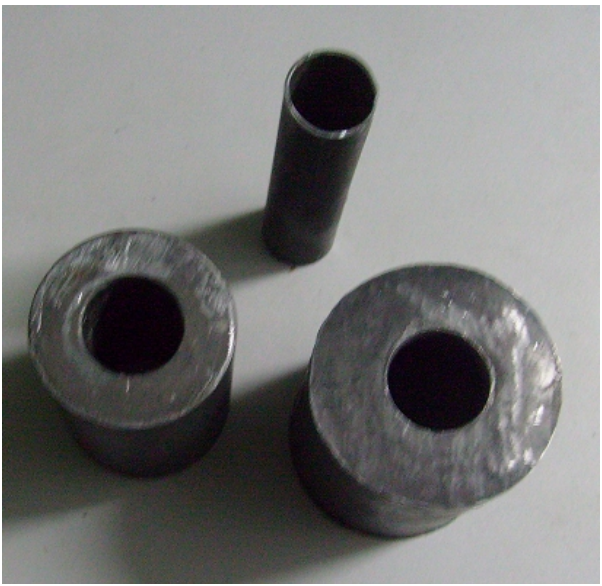

*Figure 22: Lead shielding for the muon detection.*

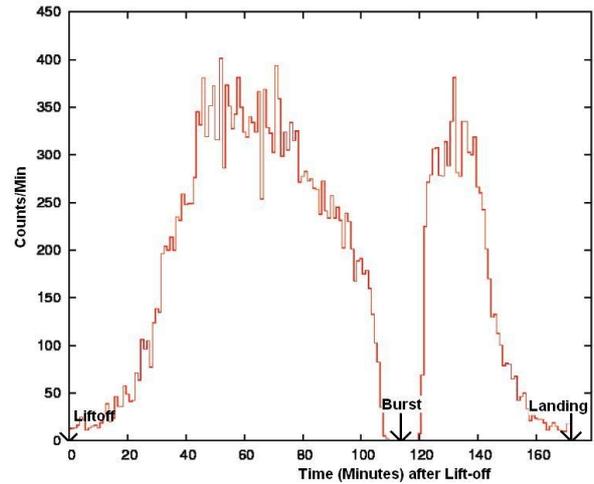

*Figure 23: Muon detection as a function of time since liftoff. This is the result of Dignity 19 mission (21 Nov. 2011). The peak occurs at around 15km, same as in cosmic ray observation, though the peak is a bit broader. There were no muon after about 30km.*

Figure 23 shows the muon count rate as a function of the height since liftoff. Note that close to the burst height there is no muon. Several missions had similar conclusions. In Fig. 24, we show another result (Dignity 17 on 11[th] November, 2011). The data obtained on the way up. Note that there is no minimum close to liftoff in both Figure 23 and Figure 24 and there is no muon emission from ground (compare Figures 4, 8 and 18 for cosmic rays).

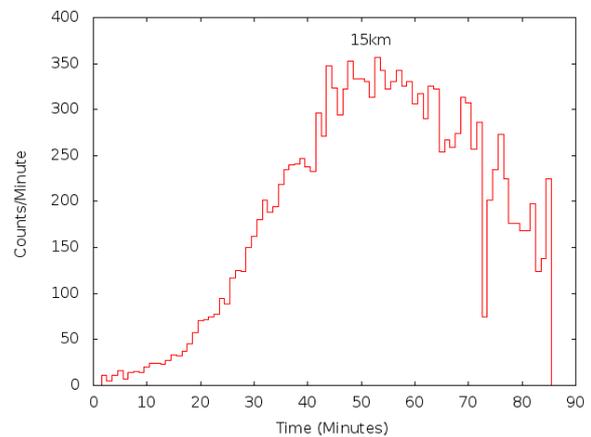

*Figure 24: Muon detection in Dignity 17 mission. The counts per minute is shown as a function of time since liftoff. The peak occurs at around 15km. Unlike in Figures 4,7, and 17, there is no minimum in Figures 22 and 23 just after liftoff and just before landing.*

**Orballoons:** In order to achieve longer duration low cost orbiter, ICSP is engaged in not only using multi-balloon concept as presented before, it is using smaller plastic balloons which could be used to more easily control the precise altitude. These non-conventional balloons have been named as Orballons as they will directly float in a given orbit without a separate booster. Figure 25 shows an example of an Orballoon with 7 micro-plastic balloons. Bursting or any one would achieve a steady orbit. The tests are in progress and the results are satisfactory.

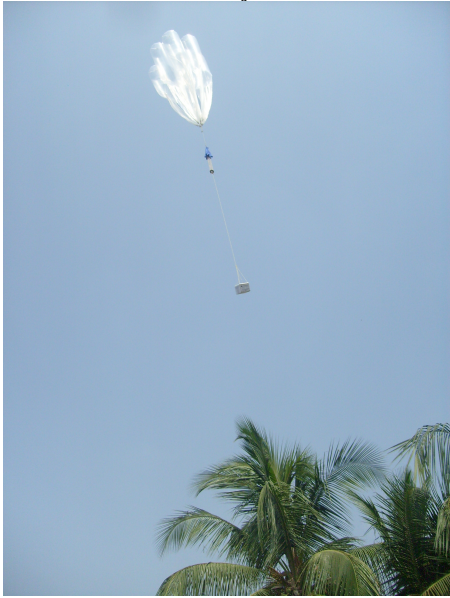

*Figure 25: Orballoon launched with seven microsized plastic balloons intended to reach a steady orbit with a single balloon burst. Tests are in progress.*

**Micrometeorite collection from space:** ICSP is collecting micro-meteorites and had two missions to this effect. Several micro-meteorites have been collected and the analysis is going on.

**All Sky Survey:** ICSP has already sent a 4" Optical telescope and brought it back with a great success. This work is further improved for all sky survey in different wavelengths.

4. **CONCLUDING REMARKS**

With miniaturizations of instruments, carrying out quality scientific experiments using low cost rubber balloons is now a reality. ICSP has been conducting pioneering work in this direction and has demonstrated that not only high quality data can be obtained, the science return per dollar is very high. All the instruments are returned reducing the cost even further. Not only that, using innovative process of multiple balloons, ICSP has achieved very long duration flights without using any conventional valve and ballast systems. Right now a Booster and Orbiter combination is launched. However, orballoons are being tested for directly floating a payload on a desired orbit.

So far, ICSP has been carrying out high energy astrophysics related studies. However, measurements of ozone, polluting chemicals, stratospheric cloud compositions, meteorites, aerosols, biological studies, are equally possible. The biggest advantage is the flexibility of launching, frequency of launching and high returns. These are most certainly extraordinary training tools for larger space missions.

One of the exciting results we presented was the detection of high energy gamma rays coming from solar flares even at the cruising heights of commercial planes. This is particularly alarming for flights during the flares and for an extended period during solar maximum. We are in a process to use GEANT4 code to simulate these events at various heights and compare with our observations.

5. **ACKNOWLEDGMENTS**

We thank the entire helping team of ICSP, especially, Hiray Roy, Uttam Sardar, Ram Chandra Das, Raj Kumar Maiti who made these experiments successful. We also thank several graduate students for accompanying us during many of the flights. We thank Dr. Dipak Debnath and Dr. Ritabrata Sarkar for helpful discussions.